\documentclass[11pt,english,12pt,onecolumn, draftcls]{IEEEtran}
\usepackage[cmex10]{amsmath}
\usepackage{times,subfigure,graphicx,epsfig,wrapfig,amsfonts,amssymb,amsthm,algorithm2e,threeparttable}
\usepackage{datetime}
\usepackage{url}
\usepackage{color}
\usepackage{newcent}
\ifCLASSINFOpdf
\usepackage{epstopdf}
\usepackage{algorithm2e}
\else

\fi
\theoremstyle{plain}

\newtheorem{lem}{Rule}

\theoremstyle{definition}

\begin{document}

\begin{titlepage}

\begin{tabular}{l        r}

\includegraphics[bb=10bp 170bp 500bp 450bp,clip,scale=0.3]{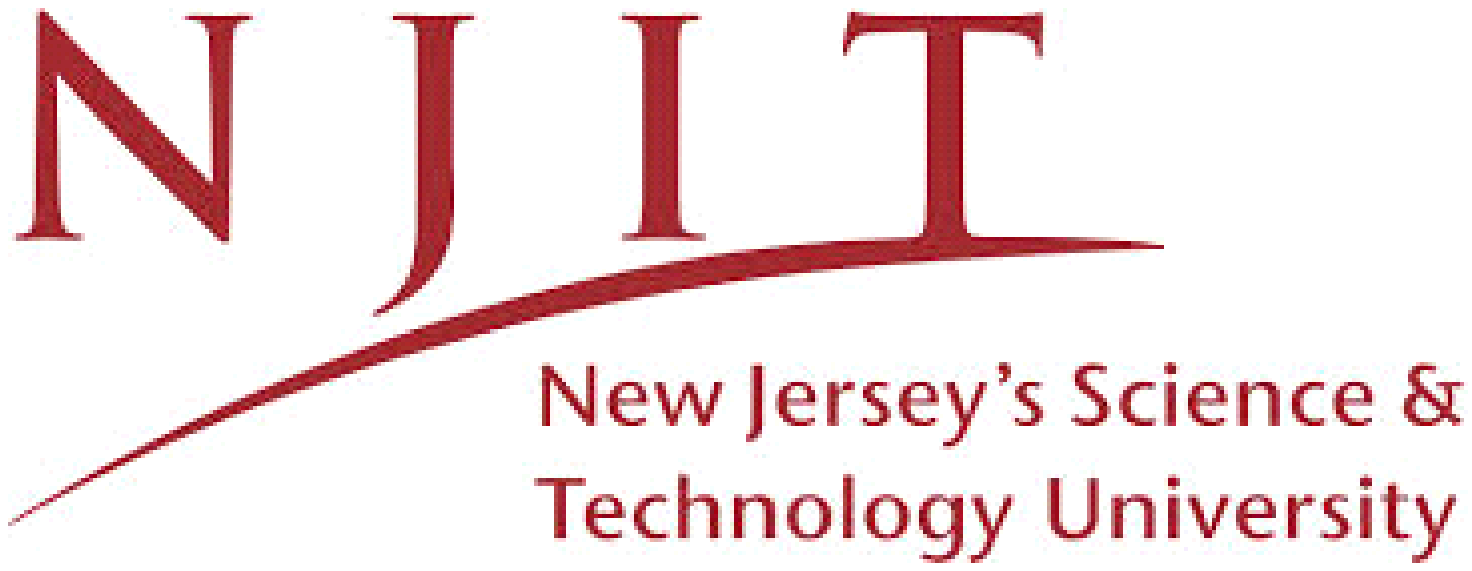} \hspace{6cm} & \includegraphics[bb=0bp 120bp 500bp 460bp,clip,scale=0.2]{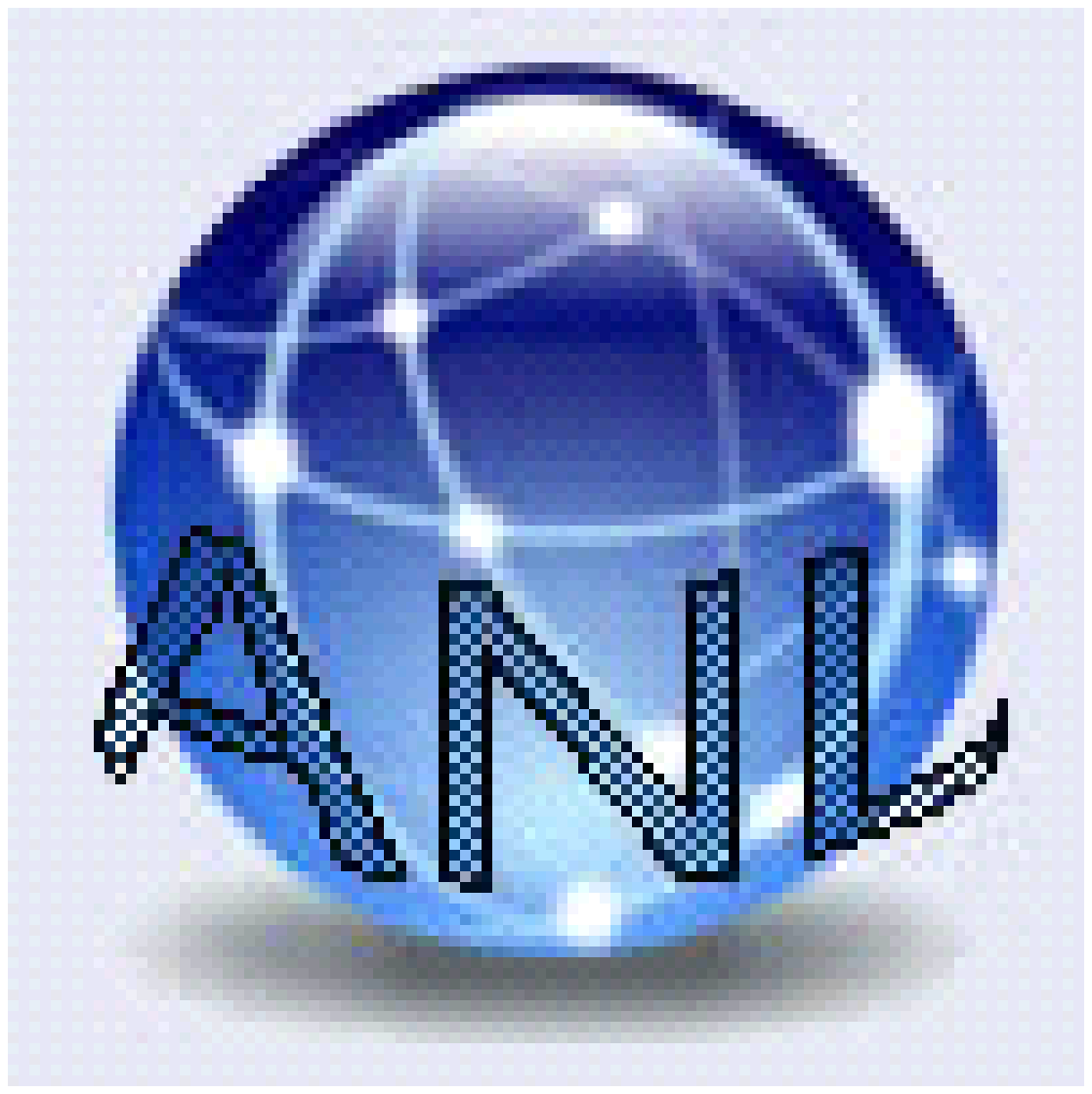}

\end{tabular}

\vspace{2cm}


\begin{center}

\textsc{\LARGE Multi-power-level Energy Saving Management for Passive Optical Networks}\\[1.5cm]

{\Large \textsc{mina taheri}}\\ 
{\Large \textsc{nirwan ansari}}\\[3cm] 



{}
{\textsc{TR-ANL-2014-001}\\
\large \usdate{March 4, 2014}}\\[3cm]

{\textsc{Advanced Networking Laboratory}}\\
{\textsc{Department of Electrical and Computer Engineering}}\\
{\textsc{New Jersy Institute of Technology}}\\[1.5cm]
\vfill

\end{center}

\end{titlepage}


\author{
\IEEEauthorblockN{ Mina Taheri Hosseinabadi and Nirwan Ansari,~\IEEEmembership{Fellow,~IEEE}\\
}

%
%

\thanks{This work was supported in part by the US National Science Foundation under Grant No. CNS-1218181.}

}


\IEEEcompsoctitleabstractindextext{%
\vspace{-1in}
\begin{abstract}
\footnote{This work is submitted to IEEE/OSA Journal of Optical Communications and Networks.}Environmental concerns have motivated network designers to further reduce energy consumption of access networks. This paper focuses on reducing energy consumption of Ethernet passive optical network (EPON) as one of the most efficient transmission technologies for broadband access. In EPON, the downstream traffic is sent from the optical line terminal (OLT) located at the central office to all optical network units (ONUs). Each ONU checks all arrival downstream packets and selects the downstream packets destined to itself. Therefore, receivers at ONUs have to always stay awake, thus consuming a large amount of energy. Contrariwise, an ONU transmitter can be triggered by the arrival of the upstream traffic and so it can go to the low power mode when no traffic is observed. Putting ONUs into the low power mode during light traffic is a known strategy for energy saving. In this article, we address the downstream challenge and also improve the ONU transmitter sleep time by proposing a simple sleep control scheme. We also propose an upstream and a downstream sleep-aware traffic scheduling scheme to avoid missing the packets during the sleep states. The proposed sleep control scheme does not need the handshake process and is based on the mutual inference at OLT and ONU. Simulation results show that the proposed scheme can save energy as much as $60\%$ when the network traffic is light.

\end{abstract}

}

\IEEEdisplaynotcompsoctitleabstractindextext

\IEEEpeerreviewmaketitle

\vspace{-0.1in}
\section{Introduction}
\IEEEPARstart{E}{}nergy consumption in data communication networks has been increasing due to an ever increase of broadband users \cite{Book13}.
In comparison with various access networks including WiMAX, FTTN (Fiber To The Node), and point to point optical access networks, passive optical networks (PONs) achieve the smallest energy consumption per transmission bit \cite{LanOnt08, ZhaUti10, zhang2010energy, zhang2011designing}. 10G Ethernet Passive Optical Network (10G-EPON) as a simple and scalable access network, which provides full-service access with high data rate for the end users, still consumes a large amount of energy in the electrical and optical devices. Environmental and economic concerns of worldwide deployment of EPON attracts the global attention to further reduce energy consumption of EPON.

Multi-power mode devices are able to reduce the energy consumption of the network by disabling certain functions \cite{Gupta2007}. EPON system adopts a TDM (Time Division Multiplex) mechanism to multiplex traffic onto a single wavelength in both upstream and downstream directions \cite{Luo2007}. In order to avoid service degradation in EPON, proper design of a MAC-layer control and scheduling scheme is required. Putting ONUs in the sleep mode during the idle time has been widely proposed in the literature.

Yan \emph{et al.} \cite{yan2010energy} proposed two scheduling algorithms to save the energy in EPON. In the first algorithm, upstream centric scheduling, the OLT buffers and sends the downstream traffic to a specific ONU during the pre-scheduled time for the ONU upstream traffic. In the second scheme, which is called downstream centric scheme (DCS), the proper time slots are scheduled for both upstream and downstream traffic. This scheme is suitable for delay sensitive applications with lower energy saving. The proposed management mechanism can save $10\%$ energy as compared to the base model with no energy saving management.

Kubo \emph{et al.} \cite{KubStu10} proposed a sleep control scheme by using adaptive link rate (ALR) control functions to reduce the ONU active time and extend the sleep periods. In the proposed sleep and periodic wake up (SPW) operation, using three way handshakes, an ONU goes to the sleep mode. The sleeping ONU wakes up periodically to check if it has any upstream or downstream traffic to send or receive, respectively. According to the monitored traffic,  the proper optical link rate is set in the multirate ONU.

In the majority of sleep control schemes, two power consumption levels are defined for an ONU. Low power consumption is defined during the sleep state in which the whole ONU goes to sleep to save the energy. High power consumption is defined in the active state in which all the sub systems of the ONU are completely functional.

Zhang and Ansari \cite{ZhaTow11} proposed to put different components to sleep under different conditions and to form a multi-power-level ONU. They did not describe state transitions between all power levels, and left some rooms for future investigations.

Modeling each ONU data transmission with an M/G/1 queue with vacation, Dhaini \emph{et al.} \cite{Dhaini11} computed the maximum ONU sleep time without proposing any sleep control scheme. However, they did not consider the downstream data transmission which can interrupt the ONU sleep time.

In this piece of work, we proposes an efficient sleep control scheme to put the ONU in different levels of energy consumption depending on both the downstream and upstream traffic situations, and thus as much as $65\%$ of the ONU energy can be saved. We investigate the trade-off between energy saving and traffic quality of service (QoS) in the simulation results. The proposed scheme is completely compatible with the MPCP control protocol and EPON standards (IEEE 802.3ah and IEEE 802.3av).

The rest of the paper is organized as follows. Section \ref{sec: Model} details the proposed sleep control scheme. Section \ref{sec:Schedul} describes the traffic scheduling algorithm. The ONU transmitter is analytically modeled, and the duration of the transmitter sleep time is calculated in Section \ref{sec:Ana}. Simulation results are given in Section \ref{sec: sim}. Section \ref{sec:Con} concludes our paper.

\vspace{-0.1in}
\section{Sleep Control scheme}
\label{sec: Model}
Two optical devices that consume the most energy in the ONU module are the optical receiver (Rx) and the optical transmitter (Tx). Owing to the burstiness of upstream and downstream traffic of the ONU, putting Rx and Tx in the sleep mode when no traffic is observed can save a significant amount of ONU energy. In some applications, e.g., file downloading, ONU has only downstream traffic. In this case, the Tx, which does not perform any task, can go to the sleep mode \cite{Mina2013}. Similarly, during the upstream transmission only, the ONU can disable the functionality of its receiver. Frequent changes of the ONU state from the active mode to the sleep mode and vice versa waste a considerable amount of energy. In order to avoid frequent changes of the states, and owing to the burstiness of the arrival traffic, it is more efficient for the ONU to spend some time in the listening state before putting each component to the sleep mode. Therefore, the ONU can immediately return to the active state upon receiving a packet.

\subsection{Multi-level Power Consumption}
\begin{figure}
\centering
\includegraphics[bb=20bp 250bp 720bp 640bp,clip,scale=0.5]{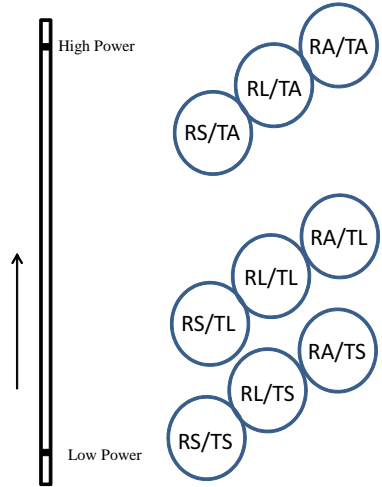}
\caption{ONU multi-level power consumption}
\label{fig:Plevel}
\vspace{-.2in}
\end{figure}

In the proposed energy saving model, there are 9 states that the ONU can enter in different situations. The ONU energy cosumption in each state is different. Fig. \ref{fig:Plevel} illustrates the level of power consumption of each state. TA, TL, and TS represent the active, listening, and sleep mode of the ONU transmitter, respectively. Likewise, RA, RL, and RS indicates the active, listening, and sleep mode of the ONU receiver, respectively. The ONU receiver typically consumes low power in the active mode \cite{Wongslp09}. Therefore, the ONU in the RL/TA and RS/TA state still consumes a large amount of energy as the transmitter is active. In the RL state, the ONU receiver is still active, but it does not receive any traffic. Thus, the power consumption is slightly lower than the active state. The ONU transmitter consumes the most portion of the ONU power. Therefore, transmitter sleep control plays a key role in reducing the ONU power consumption. 


\begin{figure*}
\centering
\includegraphics[bb=0bp 150bp 500bp 620bp,clip,scale=0.5]{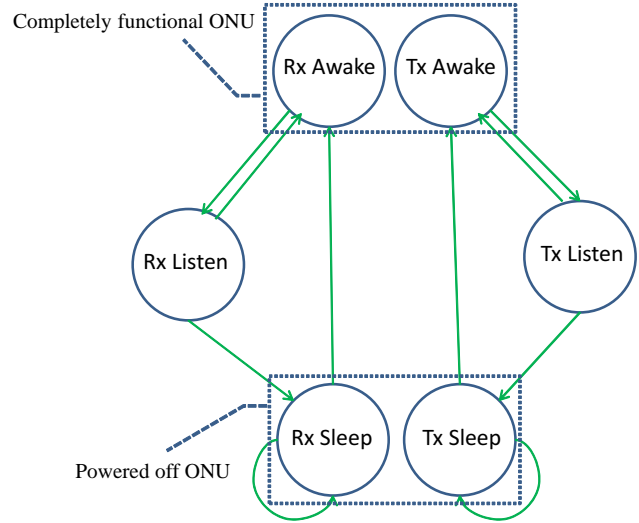}
\caption{ONU state transition.}
\label{fig:State}
\end{figure*}

To elicit the model architecture and the sleep control mechanism, we consider two different Markov models for the ONU transmitter and the receiver. The ONU can thus assume  six states: ``Rx Awake'', ``Rx Listen'', ``Rx Sleep'', ``Tx Awake'',``Tx Listen'', and ``Tx Sleep''  as shown in Fig. \ref{fig:State}.

In the ``Rx Awake'' and ``Tx Awake'' states, the ONU receiver and transmitter are fully functional, respectively. The ONU stays for a specific amount of time in the ``Rx Listen'' state before being transferred to the ``Rx Sleep''. During the ``Rx Sleep'', the ONU receiver is in the low power mode and cannot receive any traffic. 
``Tx listen'' represents the state in which the ONU transmitter is in the active mode, but it does not transmit any traffic. The power consumption of the ONU in this state is higher than the ``Tx Sleep'' state. The ONU is powered off completely whenever both transmitter and receiver are in the ``Tx Sleep'' and ``Rx sleep'', respectively, in which case the ONU consumes the least amount of power.

\subsection{Sleep control and ONU state transition}
\label{state}
The proposed sleep control scheme is based on the mutual inference between the OLT and the ONU. During the time that the receiver is in the ``Rx Awake'' state, and the transmitter is in the ``Tx Awake'', the ONU receives downstream packets and transmits upstream packets at the same time. Transition from the ``Rx Awake'' state to the ``Rx Listen'' state, and transition from the ``Rx Listen'' state to the ``Rx Sleep'' state are defined in Rule \ref{Lemma:1} and Rule \ref{Lemma:2}, respectively.

\begin{lem}
\label{Lemma:1}
If the ONU does not receive any packets within the current traffic scheduling cycle, the ONU infers that the downstream queue is empty, and the receiver enters the ``Rx Listen'' state.
\end{lem}

The OLT is aware of this transition, but the functionality of a listening ONU is the same as an active receiver from the OLT point of view.
Whenever the OLT receives the traffic destined to a listening ONU, it sends the traffic immediately. By receiving any packets from the OLT side, a listening receiver changes its state to active.

\begin{lem}
\label{Lemma:2}
If the OLT does not assign any traffic to the ONU during the pre-determined listening cycles, the ONU receiver enters the ``Rx Sleep'' state and spends a pre-determined period of time in the sleep state.
\end{lem}

Since the OLT did not allocate any traffic to the ONU in the receiver listening period, it infers that the ONU enters the sleep mode. Time duration of ``Rx Listen'' and ``Rx Sleep'' of the ONU are predefined and known for the two parties.

For the period of the ``Rx Sleep'' time, the OLT buffers the traffic destined to an asleep ONU. After the completion of the sleep period, the OLT starts sending the buffered traffic to the ONU and the ONU receiver state is changed to active. If no traffic has been buffered for the sleep ONU, the ONU skips the listening state, and returns to the ``Rx Sleep'' state. Since the OLT knows the duration of the sleep and listen states, it infers the states of the ONU, thus eliminating the necessity of handshake messages.

Transitions between listening and sleep states of the ONU transmitter are the same as the ONU receiver.  Transition from the ``Tx Awake'' state to the ``Tx Listen'' state, and transition from the ``Tx Listen'' state to the ``Tx Sleep'' state are defined in Rule \ref{Lemma:3} and Rule \ref{Lemma:4}, respectively.

\begin{lem}
\label{Lemma:3}
If the ONU does not have any traffic from the user side to send, it enters the ``Tx Listen'' state.
\end{lem}

Since the ONU bandwidth request is zero, the OLT infers that the upstream queue is empty, and is aware of the ONU change of the state. During the ONU listening time, whenever the ONU receives the traffic from the user side, it requests the proper bandwidth and change its current state to the active state.

\begin{lem}
\label{Lemma:4}
If the ONU does not receive any traffic from the user side for the pre-determined transmitter listening period, the ONU transmitter goes to the ``Tx Sleep'' state.
\end{lem}

Since the ONU transmitter can be triggered by the incoming packets from the users, the time duration of the ONU sleep transmitter varies in each ``Tx Sleep'' state. Each incoming packet can tolerate a specific amount of time in the queue before being sent. Depending on the packet class of service, packets can be buffered in the transmitter for a limited time, and transmitter in the sleep mode can sleep for a longer time. The calculation of the ONU transmitter time will be discussed in Section \ref{sec:Ana}.

The minimum energy consumption is during the time when both the ONU transmitter and receiver are in their sleep states. In our previous work \cite{Mina2013}, we described the semi-Markov chain model for the ONU receiver. We previously analyzed the delay and energy saving performance of the ONU in the situation of just having downstream traffic.  Since the Markov model for the transmitter and the receiver are equivalent, we discuss more about the traffic scheduling of the ONU in the existence of bidirectional traffic in the following section.
%

\section{Traffic scheduling algorithm}
\label{sec:Schedul}
Different settings of the listening time duration and sleep time duration, which are defined in terms of multiples of the traffic scheduling cycle ($T_{cycle}$), affect the energy saving efficiency. From now, we refer the number of cycles that the ONU transmitter or receiver stays in the sleep state as sleep cycles, and the number of cycles that the ONU transmitter or receiver stays in the listening state as listening cycles. As discussed earlier, the OLT knows all the information of the sleep control scheme implemented at each ONU which includes the listening time duration and sleep time duration. Whenever the OLT does not allocate any traffic to a specific ONU for a duration of the ONU's listening time, it infers that the ONU receiver is in the ``Rx Sleep'' state. Since the OLT is also aware of the ONU receiver sleep time, it buffers the arrival downstream traffic of the sleeping ONU until the ONU wakes up.

In the upstream direction, the ONU buffers arrival upstream traffic during the transmitter sleep time. When the transmitter wakes up, the ONU sends its bandwidth request for the next scheduling cycle to the OLT. The OLT receives the requests from all the ONUs, and assigns the bandwidth for each ONU based on dynamic bandwidth allocation (DBA). At the beginning of the scheduling cycle, the OLT sends the grant message, which contains the start transmission time and duration of the transmission (assigned bandwidth to the ONU), to each ONU.

\begin{algorithm}

 $T_{ps}$: Previous granted start time for the ONU transmission\;
 $T_S$: Current granted start time for the ONU transmission\;
 $Max_{cycle}$: maximum number of traffic scheduling cycles\;
 $Num_{cycle}$: number of traffic scheduling cycle\;
 $Num_{cycle}=0$\;
 The ONU keeps $T_{ps}$ for 1 cycle\;
 \While{ $Num_{cycle} < Max_{cycle}$}{
   \If{Tx sleep}{
  $T_{ps}=T_{ps}+T_{cycle}$\;
  }
 \eIf{Rx Sleep }{
 	\If{Tx Active}{
 	 $Request BW = min (BW_r , BW_{FBA})$\;
 	 The OLT allocates $Request BW$ to the ONU\;
 	 $T_s = T_{ps} + T_{cycle}$\;
 	 $T_{ps} = T_s$\;
 	 }
 	 }
 	{
 	OLT allocates the bandwidth based on DBA scheme\;
 	OLT sends the grant message including the assigned bandwidth and $T_s$\;
 	}
 	$Num_{cycle} = Num_{cycle} +1$\;
  	}
 \caption{Bandwidth allocation during Rx Sleep}

 \label{alg:BW}
\end{algorithm}

In the situation that the ONU receiver is in the sleep state while the transmitter is working, the ONU needs to receive the grant message for upstream data transmission. Algorithm \ref{alg:BW} describes the traffic scheduling in the case that the ONU receiver is in the sleep mode and the ONU transmitter is in the active mode. ``FBA bandwidth'' is the bandwidth that is calculated by the fixed bandwidth allocation (FBA) scheme which grants each ONU with a fixed time slot length. When the ONU reciever is in the sleep mode, the ONU requests the minimum bandwidth between ``FBA bandwidth'' ($BW_{FBA}$) and its required bandwidth ($BW_r$). The OLT is aware of the ONU receiver state and assigns the exact requested bandwidth to the ONU during the receiver sleep time. In order to set the start time of the ONU transmission in the next cycle, while the ONU receiver is in the sleep state, the OLT and the ONU keeps track of the previous ONU transmission start time ($T_{ps}$) to set $T_{ps}+T_{cycle}$ as the next start time ($T_s$). If the transmitter was in the sleep state and wakes up in the current state, the OLT and the ONU have to keep the information of the last start time. During the time that the ONU transmitter is in the ``Tx Sleep'' state, $T_{ps}$ is incremented by the ONU transmitter sleep time.

\section{Theoretical Analysis}
\label{sec:Ana}
In this section, we analyze our model with both Poisson upstream and downstream traffic. Downstream and upstream packet arrival rates are assumed to follow the Poisson process with rate $\lambda_D$ and $\lambda_U$, respectively. Service rate is also exponentially distributed in the downlink and uplink with the mean value of $\mu_D$ and $\mu_U$, respectively.

The probability of having $\alpha$ downstream arrival packets at the OLT and departure packets from the OLT in a traffic scheduling cycle is obtained by Equations \ref{Ra} and \ref{Rd}, respectively.

\begin{equation}
\label{Ra}
P_{DL}^a(\alpha)=e^{-\lambda_D T_{cycle}}\cdot (\lambda_D \cdot T_{cycle})^\alpha/\alpha!
\end{equation}

\begin{equation}
\label{Rd}
P_{DL}^d(\alpha)=e^{-\mu_D T_{cycle}}\cdot (\mu_D \cdot T_{cycle})^\alpha/\alpha!
\end{equation}

Similarly, in the upstream direction, the probability that $\alpha$ packets arrive at the ONU transmitter and depart from the ONU transmitter is obtained as follows.

\begin{equation}
\label{Ta}
P_{UL}^a(\alpha)=e^{-\lambda_U T_{cycle}}\cdot (\lambda_U \cdot T_{cycle})^\alpha/\alpha!
\end{equation}

\begin{equation}
\label{Td}
P_{UL}^d(\alpha)=e^{-\mu_U T_{cycle}}\cdot (\mu_U \cdot T_{cycle})^\alpha/\alpha!
\end{equation}

When an asleep ONU wakes up, it needs to recover the OLT clock and to be synchronized with the network which imposes an overhead to the system. Wong \emph{et al.} \cite{Wongslp09} proposed two ONU architectures which reduces the overhead time to tens of nanoseconds. Considering these two architectures, this overhead time is negligible as compared to the duration of a traffic scheduling cycle.

The Markov chain model for the ONU transmitter and the ONU receiver assumes the following states:
\begin{itemize}
\item ${A}_{T(i)}$ and ${A}_{R(j)}$ are the ``Awake'' states in which $i$ refers to the number of queued  upstream packets, and $j$ refers to the number of queued downstream packets.
\item ${L}_{T(i)}$ and ${L}_{R(j)}$ are the ``Tx Listen'' and the ``Rx Listen'' state, respectively, where $i$ ($j$) represents the number of cycles taken by the ONU transmitter (receiver) in the listening mode.
\item ${S}_{T(i)}$ and ${S}_{R(j)}$ refers to the status that the ONU transmitter and receiver stay asleep, respectively, where $i$ and $j$ represent the number of sleep cycles for the ONU transmitter and receiver to be asleep, respectively.
\end{itemize}
State transitions are performed based on the rules discussed in Sec. \ref{state}. State transition between the active states depends on the number of arrival and departure packets in each cycle.
\begin{itemize}
\item Transition from ${A}_{R(r)}$ to ${A}_{R(k)}$ (${A}_{T(r)}$ to ${A}_{T(k)}$) for $r>k$ happens when the number of departure packets are $r-k$ more than the number of arrival packets by the end of the traffic cycle.
\item Transition from ${A}_{R(r)}$ to ${A}_{R(k)}$ (${A}_{T(r)}$ to ${A}_{T(k)}$) for $r<k$ happens when the number of arrival packets are $r-k$ more than the number of departure packets by the end of the traffic cycle.
\item If all the queued and arrival packets could be served in one cycle, transition to ${A}_{R(0)}$ state occurs.
\end{itemize}

If no packet arrives at the OLT in the next traffic scheduling cycle and current state of the ONU receiver is ${A}_{R(0)}$, the ONU receiver state transits to ${L}_{R(1)}$; if the current state is ${L}_{R(k)}$, the ONU receiver state transits to ${L}_{R(k+1)}$. Transition from ${A}_{T(0)}$ to ${L}_{T(1)}$ and ${L}_{T(k)}$ to ${L}_{T(k+1)}$ also occurs in the same way for the ONU transmitter.

Consider the maximum number of listening cycles for the ONU transmitter (receiver) before going to the sleep state as $TxListen$ ($RxListen$). Whenever the number of $i$ ($j$) in ${L}_{T(i)}$ (${L}_{R(j)}$) for the ONU transmitter (receiver) reaches $TxListen$ ($RxListen$), transition from ${L}_{T(TxListen)}$ to ${S}_{T(1)}$ (${L}_{R(RxListen)}$ to ${S}_{R(1)}$) occurs.

By the time that the ONU receiver goes to the sleep state, it does not wake up until the number of sleep cycles reaches the maximum defined amount ($RxSleep$). Depending on the number of downstream arrival packets during the sleep period, transition from ${S}_{R(RxSleep)}$ to any of the receiver active states ${A}_{R(0)},{A}_{R(1)},...$ could happen. Upon waking up, if no traffic has been arrived during the sleep time, the ONU receiver will sleeps for another $RxSleep$ cycles.

Since the duration of the ONU transmitter sleep time is not pre-defined, each ONU transmitter sleeps for a different amount of time.

\subsection{The ONU transmitter sleep duration}
\label{tx}
Triggering the sleep ONU to wake up by arrival traffic can only be performed in the ONU transmitter side. Before the OLT sends the downstream traffic to the sleep ONU, the ONU receiver has to be waken up. Therefore, the receiver sleep cycles are predefined while the ONU transmitter sleep time can be calculated by considering the quality of service (QoS) class of each arrival packet \cite{Dhaini11}.

In the proposed sleep control scheme, the ONU receiver and transmitter stay active until no more packets remain in the queue. In the upstream transmission, we uses M/G/1 queues for three different priority classes: the constant bit rate (CBR) queue has the highest priority, the variable bit rate (VBR) queue has the second highest, and the Best Effort (BE) queue has the lowest priority. All dedicated packets to each queue are from the same QoS category \cite{union2001itu}. An acceptable delay threshold is defined for each category, implying that packet delay for each category cannot exceed the threshold.

Parameters of the proposed model are notated as follows:

$N$: Number of ONUs.

$i$ : Queue index $(i=0,1,2)$.

$j$ : ONU index $(j=0,1,2,..,(N-1))$.

$\lambda_{U(i,j)}$: Packet arrival rate of queue $i$ at ONU $j$.

$\mu_U$: Packet service rate.

$W_{i,j}$: The average waiting time of a packet from queue $i$ at ONU $j$.

$T_{i,j}$: Total delay of a packet from queue $i$ at ONU $j$.

$Delay_i$: Delay threshold of the packets from queue $i$.

$Sleep_{i,j}$: Sleep time of ONU $j$ based on queue $i$.

$TxSleep_j$: Maximum transmitter sleep time of ONU $j$.

$T_{propagation,j}$: Propagation delay between OLT and ONU $j$.

Total delay of a packet consists of queuing delay, service time (transmission delay), and propagation delay. Average service time and second moment of service time of the packets from queue $i$ are expressed as $X_i=1/\mu$ and $E[X_i^2]$, respectively. Using P-K formula \cite{bertsekas1992data}, the average waiting time for the packets belonging to queue $i$ is obtained as follows.

\begin{equation}
W_{i,j} = \frac{\sum_{i=0}^{2} \lambda_{i,j}E[X_i^2]}{2(1-\rho_{1,j}-...-\rho_{i-1,j})(1-\rho_{1,j}-...-\rho_{i,j})}
\end{equation}

where $\rho_{i,j}=\lambda_{i,j}/\mu$ is the utilization factor of queue $i$ at ONU $j$.

Total delay of a packet in the system is equal to the following:

\begin{equation}
T_{i,j} = W_{i,j} + 1/\mu + T_{propagation,j}
\end{equation}

The ONU transmitter sleep time based on queue $i$ depends on the delay threshold of the packets in the queue, and can be computed as:

\begin{equation}
Sleep_{i,j} = Delay_i - T_{i,j} - T_{cycle}
\end{equation}

The time duration between waking up the ONU transmitter and sending the upstream traffic is approximately one traffic scheduling cycle. At first, the ONU has to send the bandwidth request to the OLT, and then receives the grant message in which the time of the ONU transmission is granted for the upcoming cycle.

The ONU transmitter is triggered whenever the delay threshold of one of the queues is about to be met. Therefore, the maximum transmitter sleep time is equal to :

\begin{equation}
TxSleep_j = min(Sleep_{0,j},Sleep_{1,j},Sleep_{2,j})
\end{equation}

In each traffic scheduling cycle that the OLT does not receive any traffic from an ONU, it counts the cycle as the first listening cycle. From this point, the OLT just assigns the amount of bandwidth for the ONU report message. Since the OLT is aware of the duration of the ONU transmitter listening state, it can infer the change of the state from listening to the sleep state. Whenever the OLT receives the report message including the bandwidth request, it infers the change of the state to active, assigns the bandwidth accordingly, and sends back the grant message.

\section{Simulation and numerical results}
\label{sec: sim}
To validate the effectiveness of the proposed solution, we conducted simulations for both Poisson and non-Poison traffic. Recent studies have focused on either upstream or downstream traffic to make a sleep decision for the ONU. Here, we consider both upstream and downstream traffic concurrently by enabling separate sleep modes for the ONU transmitter and receiver.The SPW operation also imposes too much overhead on the system as making a decision is based on three way hand shaking. Therefore, we compare our results with the base model.  We used self-similar traffic with the Hurst parameter of 0.8. Owing to the space constraint, we have only compared the transmitter sleep time results for both traffic, and the rest of the performance evaluation is conducted for self-similar traffic. $20\%$ of the total traffic belongs to the CBR traffic with a fixed packet size of 70 bytes. The rest is equally divided between VBR and BE traffic with variable packet sizes uniformly distributed between 64 bytes and 1518 bytes. A 10G EPON supporting 32 ONUs operates at $10 Gb/s$ in the downstream direction, and $2.5 Gb/s$ in the upstream direction. Since having a fixed traffic scheduling cycle does not violate the EPON standards, we set the traffic scheduling cycle duration as $2ms$.

\begin{figure}
\centering
\includegraphics[bb=50bp 230bp 620bp 490bp,clip,scale=0.57]{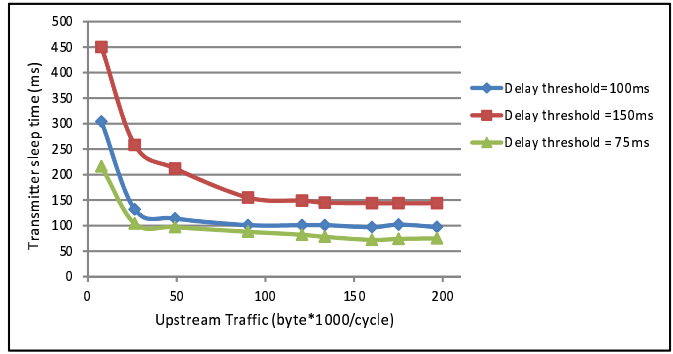}
\caption{ONU transmitter sleep time for different delay thresholds (self-similar traffic).}
\label{fig:thresh}
\vspace{-.15in}
\end{figure}

\begin{figure}
\centering
\includegraphics[bb=50bp 230bp 620bp 490bp,clip,scale=0.59]{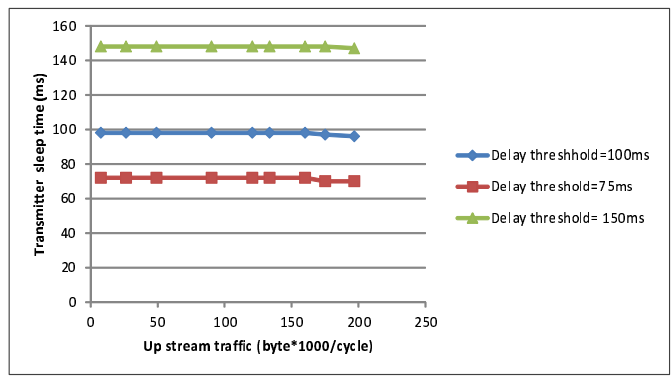}
\caption{ONU transmitter sleep time for different delay thresholds (Poisson traffic).}
\label{fig:threshn}
\vspace{-.15in}
\end{figure}

Traffic scheduling in the upstream and downstream direction is executed as follows. If the total traffic related to all the ONUs is less than the EPON capacity ($10 Gb/s$ in the downstream and $2.5 Gb/s$ in the upstream direction), all the traffic is scheduled, and all the queues become empty; otherwise, bandwidth allocation proportional to the ONU's bandwidth request is performed. The buffer size is also assumed to be infinite.

The power consumptions of the ONU transmitter and receiver in the active and sleep status are adopted from \cite{series2000transmission}. The ONU power consumptions in the ``Awake'' and ``Sleep'' state are $3.85$ W and $750$ mW, respectively. Power consumption of the ONU when the receiver is asleep and the transmitter is active would be $2.5$ W, while it decreases to $1.7$ W when the receiver is active and the transmitter goes to sleep. As for the listening status, the power consumption has to be within the range between the active and sleep power consumption. Since we cannot retrieve the exact number from publicly available literature, we performed our simulations using different values for the power consumption of the ONU receiver and transmitter in the listening status. Owing to the space constraint, we only show the results using $1.9$ W for the ``Tx listen'' state when the receiver is active and $2.8$ W for the ``Rx listen'' state when the transmitter is active, but the observations and conclusions can be similarly drawn for other values. Following Fig. \ref{fig:Plevel}, energy consumptions of the other states are estimated using the mentioned numbers.

\begin{figure}
\centering
\includegraphics[bb=90bp 250bp 620bp 480bp,clip,scale=0.7]{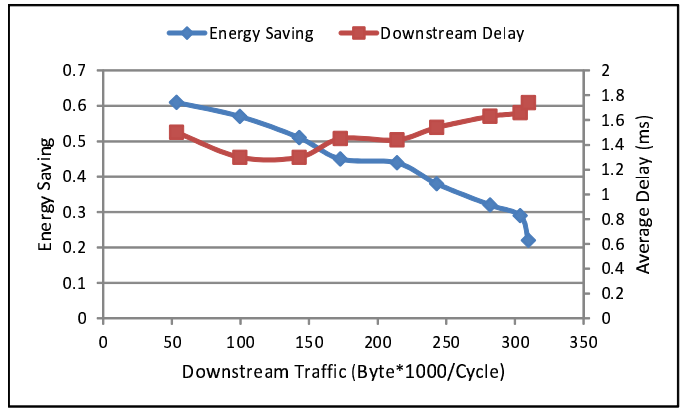}
\caption{Energy saving and delay vs. downstream arrival traffic.}
\label{fig:save}
\vspace{-.2in}
\end{figure}

Three classes of service have different delay thresholds. Figs. \ref{fig:thresh} and \ref{fig:threshn} show the direct impact of different delay thresholds on the duration of the ONU transmitter sleep time for self-similar and Poisson traffic, respectively. In each graph, we set the same delay threshold for three different classes. In the analytical results (Fig. \ref{fig:threshn}), the average transmitter sleep time cannot exceed the maximum threshold as it is calculated based on the average traffic arrival rate. However, during the low self-similar traffic load in the simulation result, the transmitter sleep time reaches the higher values when no traffic is observed. In the next simulation results, the delay threshold for CBR, VBR, and BE traffic is set as $100ms$, $1s$, and $50s$, respectively.

Fig. \ref{fig:save} illustrates the system performance for different downstream traffic loads, in which the traffic load is defined as the total arrival bytes. The number of the Rx listening cycles, Rx sleep cycles, and Tx listening cycles are the same and equal to $2$. The primary axis (left y-axis) shows the energy saving trend as compared to the base model. Increasing the traffic load decreases the probability that the ONU stays idle in the listening cycles. Therefore, the probability of the ONU entering the sleep states to save energy decreases. The secondary axis (right y-axis) indicates the imposed delay for the downstream traffic. With a low traffic load, the queuing delay is negligible, but the time to wake up from the sleep state is dominant. By increasing the traffic load, as the probability that the ONU stays in the sleep state decreases, the delay decreases. From a certain point, the queuing delay starts increasing, resulting in higher delay.

\vspace{-.15in}
\begin{figure}
\centering
\includegraphics[bb=90bp 230bp 620bp 480bp,clip,scale=0.7]{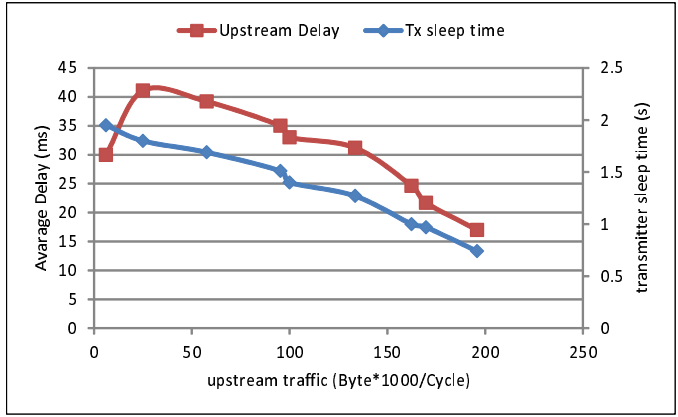}
\caption{Transmitter sleep time and upstream delay vs. upstream arrival traffic.}
\label{fig:ts-load}
\vspace{-.2in}
\end{figure}
\vspace{.2in}

From the upstream point of view, the ONU transmitter sleeps for shorter time when the network is highly loaded. Fig. \ref{fig:ts-load} represents system performance in the upstream direction. By increasing the load, the probability that the ONU transmitter goes to the sleep state decreases. Therefore, packets do not need to wait for the delay threshold to be met. The queuing delay is the dominant delay for upstream packets at high load.

\begin{figure}
\centering
\includegraphics[bb=90bp 250bp 620bp 500bp,clip,scale=0.7]{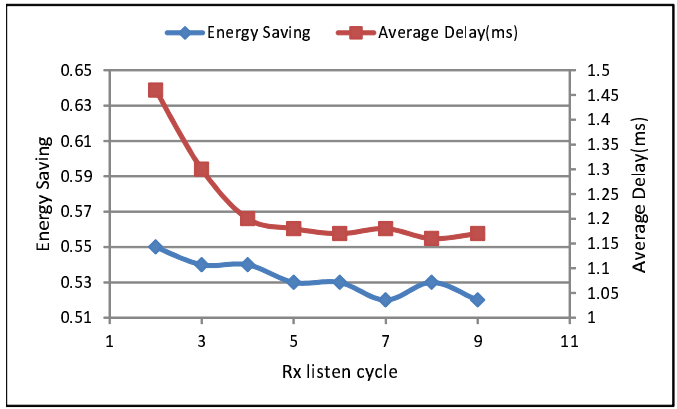}
\caption{Energy saving and downstream delay performance vs. the number of Rx listening cycles.}
\label{fig:rxl}
\vspace{-.2in}
\end{figure}

\begin{figure}
\centering
\includegraphics[bb=90bp 230bp 620bp 510bp,clip,scale=0.68]{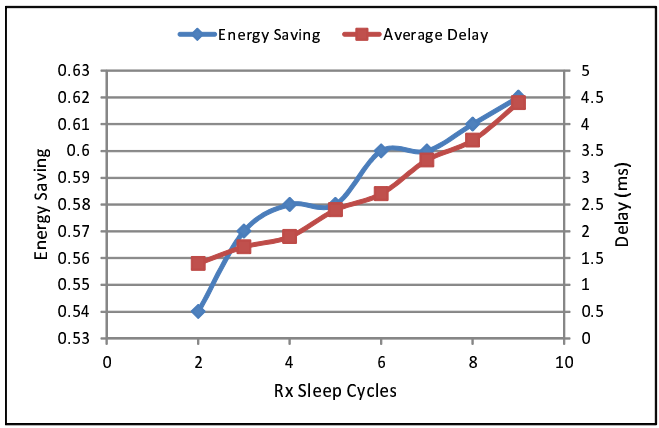}
\caption{Energy saving and downstream delay performance vs. the number of Rx sleep cycles.}
\label{fig:rxs}
\vspace{-.15in}
\end{figure}

\begin{figure}
\centering
\includegraphics[bb=80bp 230bp 620bp 510bp,clip,scale=0.68]{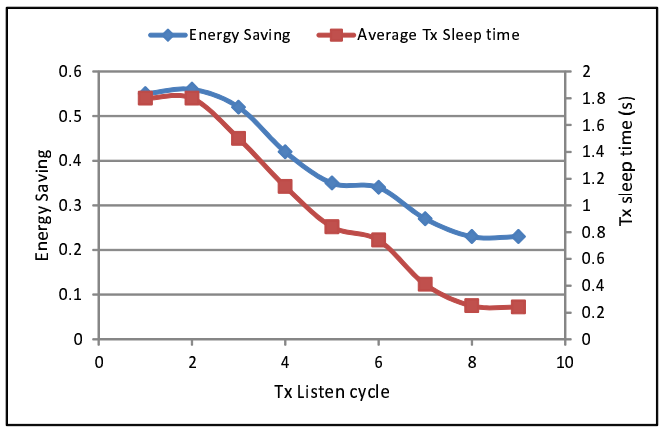}
\caption{Transmitter sleep time and upstream delay vs. the number of Tx listening cycles.}
\label{fig:txl}
\vspace{-.1in}
\end{figure}

The impact of changing Rx listening, Rx sleep, and Tx listening cycles in system performance is evaluated in Fig. \ref{fig:rxl}, Fig. \ref{fig:rxs}, and Fig. \ref{fig:txl}, respectively. As  discussed in Section \ref{sec: Model}, the ONU receiver consumes a small portion of the ONU power consumption. Therefore, changing the receiver parameters has slight effects on energy saving.  Increasing the number of listening cycles decreases the probability of staying in the ``Rx listen'' state and increases the probability of staying in the active state. Thus, the downstream traffic is  subject to lesser delay.

Increasing the number of Rx sleep cycles (Fig. \ref{fig:rxs}) results in spending more time in the low power consumption mode, thus increasing the energy saving. Since the receivers stays idle for a longer time, downstream traffic delay increases.

Changes of the ONU transmitter sleep time by increasing the number of Tx listening cycles is shown in Fig. \ref{fig:txl}. Increasing the number of listening cycles decreases the probability of keeping the ONU in the ``Rx listen'' state and increases the probability of the transmitter being active. Thus, the ONU transmitter sleep time and upstream delay both decrease by increasing the listening cycles.

\section{Conclusion}
\label{sec:Con}
We have proposed a sleep control scheme for ONUs equipped with multi-level power consumption.  Downstream traffic specifies the state of the ONU receiver, and the ONU transmitter acts according to the upstream traffic. In this paper, we have addressed the problem of putting the ONU receiver in the sleep state while the transmitter is still active. Whenever the upstream or the downstream traffic is not observed for a specific period of time, the ONU transmitter or receiver switches to the sleep mode, respectively. According to the upstream and downstream traffic arrival, the ONU decides whether to keep the transmitter and/or the receiver in the sleep mode. The proposed simple sleep control scheme and traffic scheduling algorithm are completely compatible with EPON MAC protocol. Elimination of the handshake process also makes the sleep control scheme more efficient. Our simulation results have validated the effectiveness of the proposed algorithm by saving up to $60\%$ of the ONU energy during light traffic.

\bibliographystyle{IEEETran}
\bibliography{mybib}

\end{document}